\documentclass[aps,pra,twocolumn,groupedaddress,showpacs,10pt]{revtex4-1}

\usepackage{siunitx}
\DeclareSIUnit\atoms{atoms}
\usepackage{amsmath}
\usepackage{amsfonts}
\usepackage{amssymb}
\usepackage{color}
\usepackage{graphicx}
\DeclareGraphicsExtensions{.pdf,.png,.jpg}
\graphicspath{{plot-KZ2/}, {plot-Kibble-Zurek/}}

\begin{document}

\title{Creation and counting of defects in a temperature quenched Bose-Einstein Condensate} 

\author{S. Donadello$^{1,2}$}
\author{S. Serafini$^{1}$}
\author{T. Bienaim\'e$^{1}$}
\author{F. Dalfovo$^{1,2}$}
\author{G. Lamporesi$^{1,2}$}
\email{giacomo.lamporesi@ino.it}
\author{G. Ferrari$^{1,2}$}
\affiliation{\textit{1 INO-CNR BEC Center and Dipartimento di Fisica, Universit\`a  di Trento, 38123 Povo, Italy}\\
\textit{2 Trento Institute for Fundamental Physics and Applications, INFN, 38123 Povo, Italy}}

\date{\today}

\begin{abstract}
We study the spontaneous formation of defects in the order parameter of a trapped ultracold bosonic gas while crossing the critical temperature for Bose-Einstein Condensation (BEC) at different rates. The system has the shape of an elongated ellipsoid, whose transverse width can be varied to explore dimensionality effects. For slow enough temperature quenches we find a power-law scaling of the average defect number with the quench rate, as predicted by the Kibble--Zurek mechanism. A breakdown of such a scaling is found for fast quenches, leading to a saturation of the average defect number. We suggest an explanation for this saturation in terms of the mutual interactions among defects. 
\end{abstract}

\pacs{03.75.Lm, 05.30.Jp, 67.85.De}
\maketitle

\section{Introduction\label{sec:introduction}}      % X X X X X X X X X X X X X X X X X X X X X X X X X X

A physical systems can exhibit states with different properties and symmetries depending on the values of the macroscopic parameters that describe it. Phase transitions connecting different states of a system across a critical value of a control parameter are ubiquitous in nature, from cosmology to magnetism and from classical to quantum regimes. After crossing a critical point, systems need to rearrange themselves, adapting their properties to the new conditions. While equilibrium configurations are generally well-known, the non-equilibrium dynamics of phase transitions occurring at a finite rate is far less understood. 

The Kibble--Zurek mechanism (KZM) \cite{Kibble76,Zurek85} deals with the dynamics across a phase transition involving the appearance of an order parameter in the system. The theory predicts a power-law scaling of the density of defects that the order parameter would contain after crossing the transition as a function of the quench rate. The scaling exponent depends on the intrinsic properties of the system and is the same for all systems belonging to a given universality class, independently of the microscopic details. Predictions were initially given for uniform systems undergoing a linear quench in time, and later extended to some inhomogeneous cases \cite{Zurek09,delCampo11,delCampo13}. A quantitative comparison with the observed behavior of actual systems, however, is rather challenging. For instance, the exact time at which defects are created cannot be easily estimated and very little is known in the case of nonlinear quenches or quenches where the control parameter is spatially inhomogeneous. Furthermore, interactions between defects are ignored in the KZM, whereas real systems are likely affected by such interactions during the post-quench evolution, or even at the early stages after the transition crossing.

The KZM has been experimentally observed and tested in a large variety of systems, such as liquid crystals \cite{Chuang91}, superfluid He-3 \cite{Bauerle96,Ruutu96}, thin film superconductors \cite{Carmi99,Kirtley03}, annular Josephson junctions \cite{Carmi00,Monaco06}, multiferroic crystals \cite{Chae12,Lin14} and ion chains \cite{Pyka13,Ulm13}. Ultracold atomic gases represent an ideal testbed to explore different aspects of quench mechanisms, since many of their parameters can be finely controlled and tuned. Harmonically confined gases, in oblate and prolate geometries, as well as uniform gases have been studied by quenching the temperature across the BEC transition \cite{Weiler08,LamporesiKZM13,Corman14,Chomaz15,Navon15}. In addition, quantum phase transitions at zero temperature can be crossed by varying the interaction parameters as, for instance, in the case of the transition from Mott insulator to the superfluid phase of a gas in an optical lattice \cite{Chen11,Braun15}, or from the miscible to the immiscible phase of a two-component gas \cite{Nicklas15}. 

In this article we experimentally investigate the creation of defects in a harmonically trapped ultracold gas of sodium atoms while evaporatively cooling it across the BEC transition at different rates. We extend our previous experiments \cite{LamporesiKZM13}, where we observed the KZM scaling law, by collecting further data with an improved protocol for quenching and imaging. We provide new data for different values of the transverse confinement frequencies, in order to possibly explore effects related to the dimensionality of the system. As shown in \cite{Donadello14}, the observed defects in the order parameter are quantized vortex lines which, due to the role of the transverse confinement of the elongated condensate, manifest a peculiar soliton-like character as in the solitonic-vortex structures predicted in Refs.~\cite{Brand02,Komineas03}.  

The paper is organized as follows. Section \ref{sec:theory} briefly introduces the KZM theory and focuses on the prediction of the power-law exponent in connection with dimensionality. In Sec. \ref{sec:methods}, we describe the experimental methods to quench a sodium gas across the BEC transition and explain the techniques we use to reveal the defects and characterize the system properties. In Sec. \ref{sec:results} we report on the final results and describe the different observations in case of slow or fast quenches. Conclusions are provided in Sec. \ref{sec:conclusions}.

\section{Kibble--Zurek mechanism\label{sec:theory}}      % X X X X X X X X X X X X X X X X X X X X X X X X X X

The KZM describes the defect formation in a system undergoing a continuous phase transition~\cite{Kibble76,Zurek85,delCampo14}, focussing on the spontaneous symmetry breaking that occurs at the critical value $\lambda_{\mathrm{c}}$ of a control parameter $\lambda$. If we consider the reduced control parameter $\epsilon = (\lambda_{\mathrm{c}}-\lambda)/\lambda_{\mathrm{c}}$, a second-order phase transition is characterized by the divergence of the equilibrium correlation length $\xi(\epsilon)=\xi_0/|\epsilon|^{\nu}$ and the equilibrium relaxation time $\tau(\epsilon)=\tau_0/|\epsilon|^{z\nu}$. Here $\nu$ and $z$ are the critical exponents that depend only on the universality class of the transition, while $\xi_0$ and $\tau_0$ are constants related to the specific microscopic properties of the system.

\begin{figure}[t]
\includegraphics[width=\columnwidth]{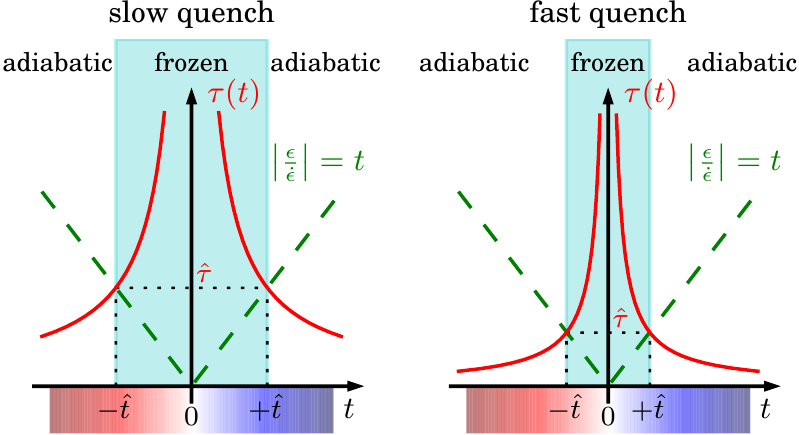}
\caption{\label{Figure1_ImpulseAdiabatic}Schematic representation of the timescales around the transition at $t=0$, where the relaxation time $\tau=\tau_0 \tau_{\mathrm{Q}}^{\nu z}/t^{\nu z}$ (solid lines) diverges. In the KZ theory, the extent of the frozen region is approximated by the freeze-out time $\hat{t}$, corresponding to the point where $\tau$ equals the time distance from the transition $|\epsilon/\dot{\epsilon}|$ (dashed lines). Qualitative differences between slow (small $\tau_{\mathrm{Q}}$) and fast (large $\tau_{\mathrm{Q}}$) quenches are illustrated.}
\end{figure}

The phase transition is crossed with a variable quench rate defined by the time derivative of the control parameter~$\dot{\epsilon}$. If we consider a quench that is linear in time, we can express the control parameter as $\epsilon(t) = t/\tau_{\mathrm{Q}}$. By doing so, the quench time $\tau_{\mathrm{Q}}=1/\dot{\epsilon}$ becomes the  relevant time scale for the quench. The relaxation time $\tau$ diverges at the critical point $t=0$ as qualitatively represented in Fig.~\ref{Figure1_ImpulseAdiabatic}. Starting from a high-symmetry state at $t \ll 0$, where $\tau$ is small, the spontaneous symmetry-breaking occurs while driving the system across the transition: $\tau$ diverges and the dynamics freezes because the system is no longer able to adiabatically follow the variation of the control parameter. The dynamics becomes adiabatic again for $t \gg 0$.

In the context of the KZM, the crossing of the continuous transition is approximately described by the presence of three distinct regimes. As illustrated in Fig.~\ref{Figure1_ImpulseAdiabatic}, the frozen regime is the one during which the relaxation time $\tau$ is larger than the time distance from the transition. The time for which the distance from the transition equals the relaxation time is called freeze-out time $\hat{t}$: the system is considered frozen for $|t|<\hat{t}$ and adiabatic elsewhere. By introducing $\hat{\epsilon}$ as $\epsilon(\hat{t})$, one can express the relaxation time at $\hat{t}$ as
\[
\hat{\tau}=\tau(\hat{\epsilon})=\frac{\tau_0}{|\hat{\epsilon}|^{\nu z}}=\frac{\tau_0\tau_{\mathrm{Q}}^{\nu z}}{|\hat{t}|^{\nu z}}\,.
\]
From the definition $\tau(\hat{\epsilon})=|\hat{t}|$, the freeze-out time results to be
\begin{equation}
\hat{t} = \left(\tau_0\tau_{\mathrm{Q}}^{z\nu}\right)^\frac{1}{1+z\nu}\,.
\label{eq:kzm-t-hat}
\end{equation}

As a consequence of causality and of the frozen dynamics, spatially disconnected regions of the system can independently choose different values for the order parameter while crossing the transition. The KZM predicts that the average size of such domains $\hat{\xi}$ is the correlation length at $\hat{\epsilon}$,
\begin{equation}
\hat{\xi} = \xi(\hat{\epsilon}) = \xi_0\left(\frac{\tau_{\mathrm{Q}}}{\tau_0}\right)^\frac{\nu}{1+z\nu}\,.
\label{eq:kzm-xi-hat}
\end{equation}

Since topological defects can arise from discontinuities of the order parameter, the presence of independent domains after the freeze-out can result in the formation of defects at the boundaries between domains. The average density of defects $n$ in the system can be calculated \cite{delCampo14} as the ratio between the size of the defects $\hat{\xi}^d$ and the size of the domains $\hat{\xi}^D$, with $d$ and $D$ the dimensionality of the defects and of the space, respectively:
\begin{equation}
n \sim \frac{\hat{\xi}^d}{\hat{\xi}^D} = \frac{1}{\xi_0^{D-d}} \left(\frac{\tau_0}{\tau_{\mathrm{Q}}}\right)^{(D-d)\frac{\nu}{1+z\nu}}\,.
\label{eq:kzm-defects}
\end{equation}

As discussed in Ref.~\cite{delCampo14}, in order to obtain a better estimate of $n$, $\hat{\xi}$ should be multiplied by a factor $f$, with the value of $f$ being in the range 1--10, depending on the specific model. Such a correction reflects also on the determination of $\hat{t}$.

The main prediction of the KZM is the power-law scaling of the density of defects with the quench time $n \propto \tau_{\mathrm{Q}}^{-\alpha}$. The above derivation is obtained considering a homogeneous system, and predicts a power-law exponent
\begin{equation}
\alpha_\text{homog} = (D-d)\frac{\nu}{1+z\nu}\,.
\label{eq:alpha-homog}
\end{equation}
For inhomogeneous systems the transition does not occur simultaneously everywhere, and the theory must keep the external trapping potential into account, with the introduction of local parameters. Under the assumptions of a linear quench and of a uniform control parameter in the whole system, for a harmonic trap the power-law exponent becomes~\cite{Zurek09,delCampo11}
\begin{equation}
\alpha_\text{harm} = (D-d)\frac{1+2\nu}{1+\nu z}\label{eq:alpha-harm}\,.
\end{equation}

The prediction of Eq.~\eqref{eq:kzm-defects} refers to the total density of topological defects at their creation, while the post-quench dynamics is completely ignored in the model. Following the KZM, the relevant time, that should be considered for counting defects after the transition crossing, is the freeze-out time of Eq.~\eqref{eq:kzm-t-hat}.

The power-law exponents \eqref{eq:alpha-homog} and~\eqref{eq:alpha-harm} depend on the critical exponents $\nu$ and $z$, whose values are not know {\it a priori}. A first reasonable choice for the critical exponents could come from a pure mean-field calculation, that gives $\nu=1/2$ and $z=2$. Going beyond mean-field theory, the so-called F-model~\cite{Hohenberg77} predicts $\nu=2/3$ and $z=3/2$. In this work, we will consider the values taken from the F-model, since recent experiments with ultracold gases~\cite{Donner07,Navon15,Chomaz15} seem to support this choice for the universality class of a 3D Bose-Einstein condensate.

The exponent $\alpha$ depends also on the quantity $(D-d)$, related to the dimensionality of the system and of the defect. Using superfluid gases, one generally deals with two kinds of defects: \textit{solitons} or \textit{vortices}. In the case of solitons the order parameter varies along one direction only and one has $(D-d)=1$, which includes 2D planar solitons in 3D systems, linear defects in 2D systems or point-like defects in 1D systems. In the case of quantized vortices, instead, the order parameter varies by $2\pi$ around a singular point, hence it needs two dimensions to be defined, so that $(D-d)=2$. Linear vortical filaments in 3D systems or pointlike defects in 2D systems belong to the same dimensionality class in the framework of the KZM. The values of $\alpha$ that might be interesting in this work, in relation to the formation of vortices or solitons, are reported in Table~\ref{tab:predicted-exponents} for the homogeneous and harmonic cases.

\begin{table}
\caption{Power-law exponents $\alpha$ predicted for the KZM from Eqs.~\eqref{eq:alpha-homog} and~\eqref{eq:alpha-harm} in a homogeneous gas and in a harmonically trapped gas, if the critical exponents predicted by the F-model, $\nu=2/3$, $z=3/2$~\cite{Hohenberg77} are used, and assuming the defects to be either solitons or vortices.\label{tab:predicted-exponents}}
\begin{ruledtabular}
\begin{tabular}{lccc}
 & $D-d$ & $\alpha_\text{homog}$ & $\alpha_\text{harm}$ \\
\cline{2-4}
solitons & $1$ & $1/3$ & $7/6$ \\ 
vortices  & $2$ & $2/3$ & $7/3$ \\ 
\end{tabular}
\end{ruledtabular}
\end{table}

\begin{figure}[t]
\includegraphics[width=\columnwidth]{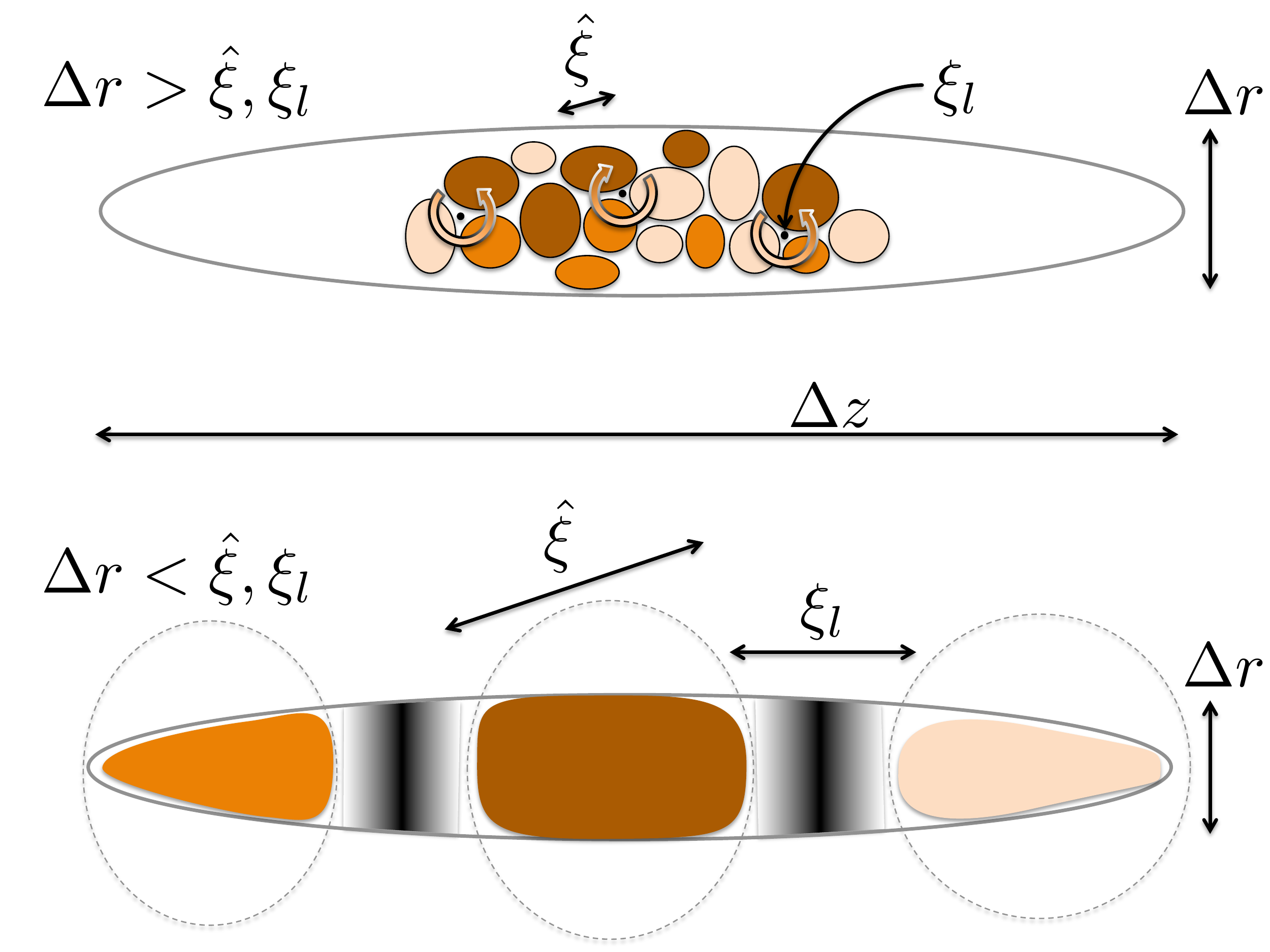}
\caption{\label{Figure2_Cartoon} Qualitative illustration of the spontaneous formation of vortices or solitons in elongated systems depending on the relative magnitude of the transverse size $\Delta r$, the domain size $\hat{\xi}$, and the healing length $\xi_l$.}
\end{figure}

The nature of the defects which form spontaneously at the transition depends mainly on three relevant length scales: the size of the system at the transition, the average domain size $\hat{\xi}$ (see Eq.~(\ref{eq:kzm-xi-hat})) and the healing length $\xi_l=1/\sqrt{8\pi a \rho}$, where $a$ is the scattering length and $\rho$ is the atomic density. Clearly the system size has to be larger than $\hat{\xi}$ and $\xi_l$ at least along one direction, otherwise different domains are not allowed and defects cannot form. Let us consider an elongated system like the one available in our laboratory, with a long axial size $\Delta z$ and a shorter transverse width $\Delta r$. As illustrated in Fig. \ref{Figure2_Cartoon}, if $\Delta r$ is simultaneously larger than $\hat{\xi}$ and $\xi_l$, then vortices can be defined and they will likely be the most probable defect forming with the quench. If, instead, $\Delta r$ is smaller or of the same order of magnitude of at least one among $\hat{\xi}$ and $\xi_l$, then the domains are forced to line up one next to the other along the long axis of the system and soliton formation is favored.

However, what is formed at the transition does not necessarily coincide with what is observed at the end of the quench. This depends on the dynamics and the stability of the defects. In order to quantitatively estimate the stability of different defects in a given BEC~\cite{Brand02,Mateo14,Tylutki15}, it is convenient to introduce the dimensionless parameter
\begin{equation}
\label{eq:gamma}
\gamma = \frac{R_\text{rad}}{2\xi_l} = \frac{\mu}{\hbar \omega_\text{rad}}\,,
\end{equation}
where $R_{\mathrm{rad}}$ is the transverse Thomas-Fermi radius of the formed BEC after the cooling ramp. When $\gamma$ is of the order of $1$ or smaller, the condensate is too narrow to host a vortex, while solitons (planar defects perpendicular to the long axis of the system) are allowed and, in this limit, they are also stable. When $\gamma \gg 1$,  $\xi_l$ is smaller than the system size along any direction. In principle both planar solitons ($d=2$) and vortex filaments ($d=1$) can exist. However, the larger is $\gamma$, the higher is the probability for solitons to decay via snake instability \cite{Muryshev99} into vortical lines with lower energy \cite{Mateo14}. Hence for $\gamma \gg 1$, if solitons are initially formed via the KZM, they likely decay into vortices.

\section{Methods\label{sec:methods}}    % X X X X X X X X X X X X X X X X X X X X X X X X X X

\subsection{BEC production \label{sec:BEC-production}}

We produce ultracold samples of sodium atoms in the internal state $|F,m_{\mathrm{F}}\rangle=|1,-1 \rangle$ in a cigar-shaped harmonic magnetic trap. A detailed description of the experimental apparatus can be found in Ref.~\cite{Lamporesi13}. The aspect ratio of the trap is defined as $AR = \omega_\text{rad}/\omega_\text{ax}$,  with $\omega_\text{rad}/2\pi$ and $\omega_\text{ax}/2\pi$ being the trapping frequencies along the radial and axial directions. In the experiment we keep $\omega_\text{ax}$ fixed to 2$\pi\times 13$ Hz and vary $\omega_\text{rad}$ from 2$\pi \times76$ Hz to 2$\pi\times 214$ Hz, hence exploring a variation of $AR$ from $5.8$ to $16.5$. The radius $R_\text{rad}$ varies consequently by a factor $2$, from about $20$ to $10$ $\mu$m.

The sample is cooled down via forced evaporative cooling and pure BECs of typically $10^7$ atoms are produced. The part of the evaporation ramp in the vicinity of the transition -- from now on called quench ramp -- is performed at different rates, from  $50$~kHz/s to $2$~MHz/s, in order to explore the KZM with the appearance of a few or many topological defects in the condensate. For a given $AR$, the variation of the final $R_\text{rad}$ at the end of the evaporation ramp for different $\tau_{\mathrm{Q}}$ is negligible.

The quench ramp is followed by a variable wait time $t_{\mathrm{w}}$, during which an RF shield is kept on to prevent from heating. After that, the atoms are released from the trap and observed with a triaxial absorption imaging, as in Ref. \cite{Donadello14}.

\subsection{Determination of $\tau_{\mathrm{Q}}$}   % X X X X X X X X X X X X X X X  

\begin{figure}[t!]
\includegraphics[width=1\columnwidth]{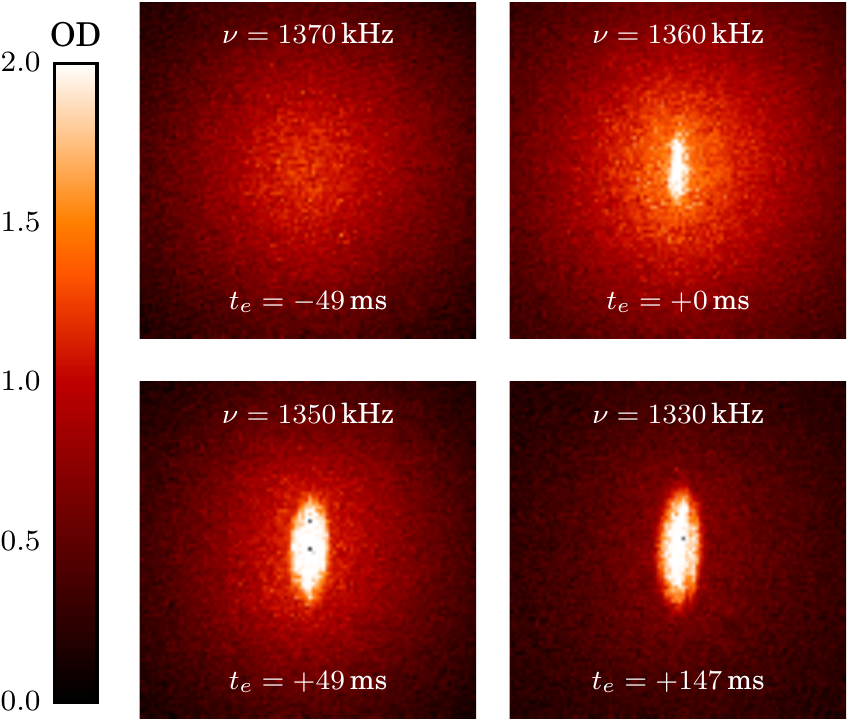}
\caption{\label{Figure3_BECTransition} Sequence of experimental absorption pictures of the atomic sample around the transition, occurring at $\nu_{\mathrm{c}} \sim$ 1360 kHz for a ramp of 203 kHz/s with AR = 10.1. All these pictures have been taken after a time of flight of 50 ms. At $t=0$ a small condensate fraction of $\sim 1\%$ of the atoms appears in the thermal cloud and grows for $t>0$. In the last picture the condensate appears much more definite, and a defect is clearly visible in it. With such a technique it is almost impossible to detect the presence of defects around $t\sim 0$. They start to become visible about 100 ms after the transition.}
\end{figure}

As discussed in Sec. \ref{sec:theory}, a finite cooling rate across the transition makes the system freeze for a variable amount of time ($\simeq 2\hat{t}$). The defect number depends on the system and on the quench parameters within such a time interval. However, neither the exact time of creation nor the time at which they start to interact are precisely known from theory. One could argue that defects are created at $-\hat{t}$, when the system starts freezing, or at $+\hat{t}$, when the system is again able to follow the changes of the external control parameter after breaking the symmetry. The time $\hat{t}$ itself is also hard to estimate (see Sec. \ref{sec:theory}).

\begin{figure}[b!]
\includegraphics[width=1\columnwidth]{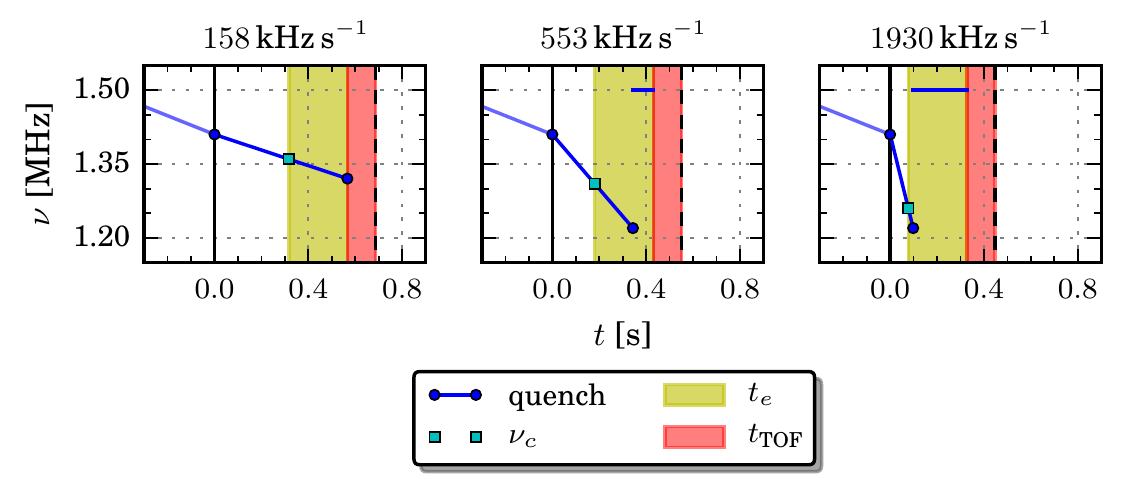}
\caption{\label{Figure4_Ramps} Experimental quench sequence for three ramp speed values. The radio frequency $\nu$ (solid blue lines) is plotted as a function of time relative to the start of the final quench ramp. Negative times refer to the preliminary evaporative cooling stage above $T_c$, which is the same for all samples. The critical point $\nu_{\mathrm{c}}$ of the BEC transition is reported for each quench ramp (cyan squares). The evolution time $t_{\mathrm{e}}$ is kept constant, relative to the transition (shadowed in yellow). After a time $t_{\mathrm{e}}=250$ ms the sample is released, let expand for a fixed TOF of $120$~ms (shadowed in red), and finally observed with absorption imaging (dashed lines). For faster quenches a waiting time is added while keeping a constant RF shield on. The quench time $\tau_{\mathrm{Q}}$ in the three cases is 970 ms, 240 ms and 60 ms.}
\end{figure}

The impossibility to unambiguously identify the time of defect creation and of interacting dynamics, combined with the observation of a finite lifetime for the stochastic defects~\cite{Donadello14,Serafini15}, suggests us to measure the defect number after a fixed given evolution time $t_{\mathrm{e}}$ from the transition point, which is clearly identifiable as can be seen in Fig. \ref{Figure3_BECTransition}. This protocol differs from the one used in Ref.~\cite{LamporesiKZM13}, where the quench rate was varied while keeping initial and final temperatures fixed for all different ramps. In addition, we are in the condition to precisely identify the critical radio-frequency $\nu_{\mathrm{c}}$ and temperature $T_{\mathrm{c}}$ where the BEC transition occurs for any given experimental condition, i.e., for each choice of $\omega_{\mathrm{rad}}$ and quench ramp. 

\begin{figure}[t!]
\includegraphics[width=\columnwidth]{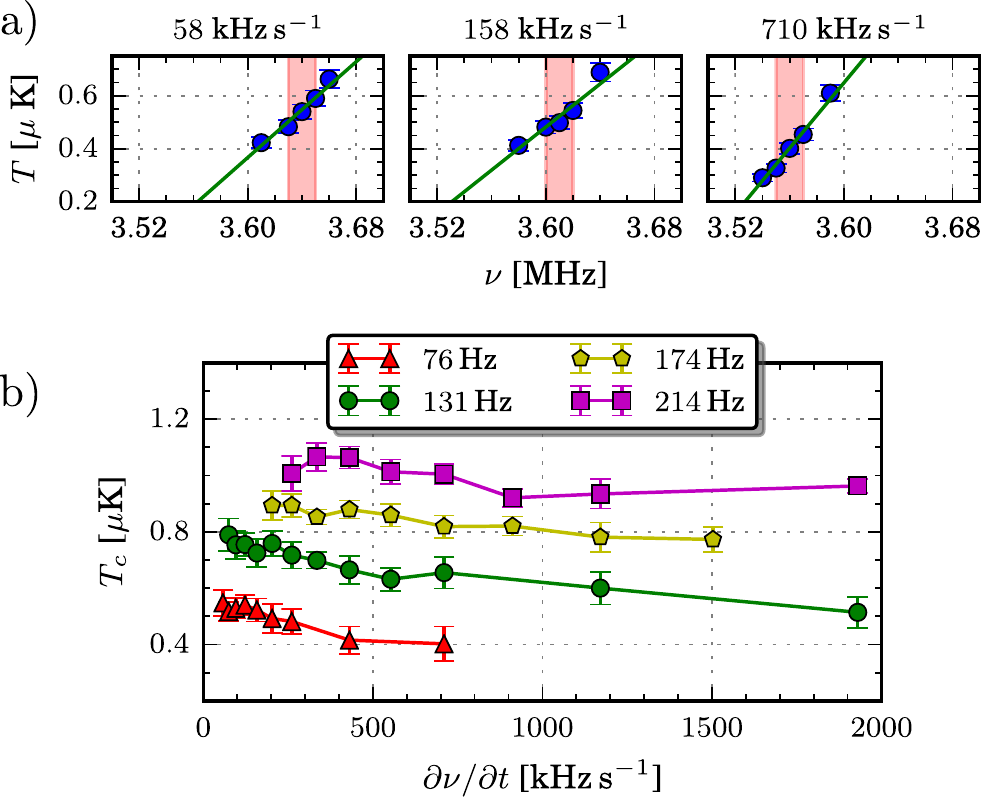}
\caption{\label{Figure5_Tc} (a) Measured temperature across $T_{\mathrm{c}}$ for a slow (58 kHz/s), medium (158 kHz/s) and fast (710 kHz/s) quench ramp. A linear fit is shown and the transition region is highlighted. (b) Critical temperature at the BEC transitions as a function of the ramp speed for different $\omega_{\mathrm{rad}}$. These values of $T_{\mathrm{c}}$ are used in Eq.~\eqref{eq:tau-Q} for the determination of the quench time.}
\end{figure}

Since defects need a time of the order of a hundred ms to become clearly detectable in terms of density depletion, we choose a time interval $t_{\mathrm{e}}$ from the transition ($t_{\mathrm{e}}>$ 100 ms), after which the atoms are released from the trap. The time $t_{\mathrm{e}}$ is kept fixed for any given quench ramp.  If $t_{\mathrm{e}}$ is reached before the end of the quench ramp (slowest ramps), the ramp is interrupted before its completion, the atoms are released from the trap and observed after a long time of flight (TOF);  a larger fraction of the atoms will remain in the non-condensed state with a higher temperature. Else, if $t_{\mathrm{e}}$ is longer than the evaporation ramp (fastest ramps) a waiting time $t_{\mathrm{w}}$ is added in the sequence after the end of the ramp and before the observation. Figure \ref{Figure4_Ramps} shows the relevant timescales in case of ramps with different rates.

The radio frequency ramp causes a temperature quench. We have verified that, for all samples, the evaporation ramp is always slow enough for the system temperature to adiabatically follow the variation of the radio-frequency. The reduced control parameter can thus be expressed as $\epsilon = 1-(T/T_{\mathrm{c}})$ and the quench rate as $\dot{\epsilon}=-(1/T_{\mathrm{c}}) (\partial T/\partial t)$. The temperature variation in time, $(\partial T/\partial t)$, is indirectly controlled {\it via} the speed of the evaporation ramp $(\partial \nu/\partial t)$. In Fig.~\ref{Figure5_Tc}a we show the measured temperature as a function of frequency around $T_{\mathrm{c}}$ for three ramp slopes. With a linear fit to each dataset we extract $(\partial T/\partial \nu)$, that is roughly constant in the different experimental conditions and has an average value of $(\partial T/\partial \nu)=\SI{4.5\pm0.9}{\nano\kelvin\per\kilo\hertz}$. Following such a procedure and referring to the quench time introduced in Sec.~\ref{sec:theory}, we have all the ingredients to estimate $\tau_{\mathrm{Q}}$, for any ramp and experimental condition, as
\begin{equation}
\label{eq:tau-Q}
\tau_{\mathrm{Q}} = -T_{\mathrm{c}} \left( \frac{\partial T}{\partial t} \right)^{-1}  = - T_{\mathrm{c}} 
\left( \frac{\partial T}{\partial \nu} \frac{\partial \nu}{\partial t} \right)^{-1} \, ,
\end{equation}
where $T_{\mathrm{c}}$ is identified by the onset of the BEC as in Fig.~\ref{Figure3_BECTransition}, and reported in Fig.~\ref{Figure5_Tc}b as a function of $(\partial\nu/\partial t)$. This procedure improves the determination of $\tau_{\mathrm{Q}}$ compared to our previous work \cite{LamporesiKZM13}, where $\tau_{\mathrm{Q}}$ was simply defined as the quench time duration.

\subsection{Defect observation and counting}

The total atom number at the transition is kept fixed to $N_{\mathrm{c}} = \SI{27 \pm 1 e6}{\atoms}$, by tuning the number of atoms involved in the early stage of laser cooling. We observe that, even if the number of atoms at the transition is almost constant, the final number of atoms in the BEC varies significantly with $\tau_{\mathrm{Q}}$, both because of different cooling efficiency and of the finite evolution time. 

The natural size of the defects in the trapped system, at the end of the cooling ramp, is of the order of the \textit{in-situ} healing length $\xi_{\mathrm{l}}$, which, at the end of the cooling ramp, is as small as $100$-$200$~nm. Therefore we let the BEC expand for a long TOF ranging from $80$~ms to $150$~ms, depending on the trap $AR$ used. In such a way the defects become larger than our imaging resolution of $3\ \mu$m and acquire the specific twisted shape described in Ref. \cite{Donadello14}.
The presence of a levitating magnetic field gradient makes it possible to achieve such a long TOF preventing the BEC from falling. 
% During the long expansion the elongated BEC inverts its aspect ratio and acquires a pancake shape.

The measured defect number $N_{\mathrm{d}}$ is averaged over many experimental realizations in order to get good statistical samples for each experimental condition: due to the power-law scaling of the defect number in the KZM we iterate longer (typically a few tens) for bigger $\tau_{\mathrm{Q}}$, where $\langle N_{\mathrm{d}} \rangle$ is smaller. The error bars for $\langle N_{\mathrm{d}} \rangle $ are estimated as $\Delta N_{\mathrm{d}} = \sqrt{\delta N^2+(1/N)}$, that is the sum in quadrature of the standard error of the mean ($\delta N$) and of a resolution term ($1/\sqrt{N}$) decreasing with the number of observations $N$.

\section{Results\label{sec:results}}

\begin{figure}[!b]
\includegraphics[width=\columnwidth]{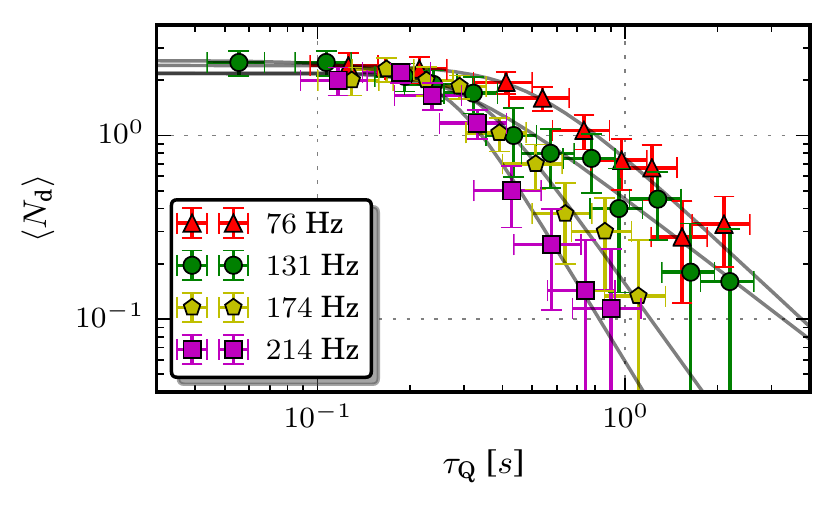}
\caption{\label{Figure6_DefectScaling} Average number of defects $\langle N_{\mathrm{d}} \rangle$ as a function of the quench time $\tau_{\mathrm{Q}}$ for several transverse confinements and with a fixed evolution time $t_{\mathrm{e}}=250$ ms. Each point with its error bar is calculated by averaging over tens of experimental realizations. A power-law behavior (linear in the log-log scale) is observed for large $\tau_{\mathrm{Q}}$, while a saturation effect is present for small $\tau_{\mathrm{Q}}$. The lines correspond to fitting functions of the form $\langle N_{\mathrm{d}} \rangle =  N_{\mathrm{sat}} [1+(\tau_{\mathrm{Q}}/\tau_{\mathrm{Q}}^0)^{2\alpha}]^{-1/2}$.}
\end{figure}

We measure the average number of defects $\langle N_{\mathrm{d}} \rangle$ as a function of the quench time $\tau_{\mathrm{Q}}$ using the quench method described in Sec.~\ref{sec:methods}. Figure~\ref{Figure6_DefectScaling} shows the results for four different transverse confinements, $\omega_{\mathrm{rad}}/2\pi=76, 131, 174$ and $214$~Hz. The data clearly exhibit two regimes: for large $\tau_{\mathrm{Q}}$, the observed $\langle N_{\mathrm{d}} \rangle$ decreases as a power law, while for small $\tau_{\mathrm{Q}}$ it saturates around a constant value $N_{\mathrm{sat}}$.  We fit the data with the function $\langle N_{\mathrm{d}} \rangle =  N_{\mathrm{sat}} [1+(\tau_{\mathrm{Q}}/\tau_{\mathrm{Q}}^0)^{2\alpha}]^{-1/2}$, which has the same behavior. The fitting parameters $\alpha$ and $\tau_{\mathrm{Q}}^0$ represent the power-law exponent and the characteristic quench time at which the two regimes interpolate, respectively. The results are reported in Table \ref{tab:fit-results}.

\begin{table}[!h]
\caption{Saturation number $N_{\mathrm{sat}}$, crossover time $\tau_{\mathrm{Q}}^0$ and power-law exponent $\alpha$ extracted by fitting the data of Fig.~\ref{Figure6_DefectScaling} for various radial trapping frequencies and aspect ratios.\label{tab:fit-results}}
\begin{ruledtabular}
\begin{tabular}{ccccc}
{$\omega_{\mathrm{rad}}$}[s$^{-1}$] &{$AR$} & {$N_{\mathrm{sat}}$} & {$\tau_{\mathrm{Q}}^0$ [\si{\second}]} & {$\alpha$} \\ 
\hline
$2\pi\times$76&\num{5.8} & \num{2.4\pm 0.3} & \num{0.49 \pm 0.13} & \num{1.6 \pm 0.4} \\
$2\pi\times$131&\num{10.1} & \num{2.6\pm 0.3} & \num{0.26 \pm 0.08} & \num{1.3 \pm 0.3} \\
$2\pi\times$174&\num{13.4} & \num{2.2\pm 0.3} & \num{0.32 \pm 0.08} & \num{2.3 \pm 0.8} \\
$2\pi\times$214&\num{16.5} & \num{2.2\pm 0.3} & \num{0.27 \pm 0.06} & \num{2.8 \pm 0.9}
\end{tabular}
\end{ruledtabular}
\end{table}

\subsection{Large $\tau_{\mathrm{Q}}$: Power-law scaling of defects \label{sec:powerlaw}}

\begin{figure}[b!]
\includegraphics[width=\columnwidth]{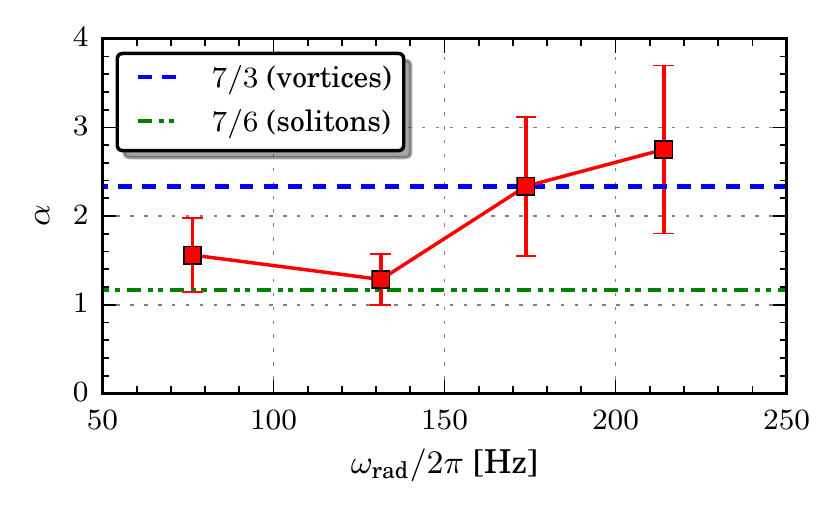}%
\caption{\label{Figure6_alpha} Power-law exponent $\alpha$ obtained by fitting the data of Fig.~\ref{Figure6_DefectScaling} for different radial frequencies. The KZ predictions for solitons and vortices in an harmonically trapped 3D condensate, from Table~\ref{tab:predicted-exponents}, are shown as horizontal lines for comparison. }
\end{figure}

In case of small cooling rates, the average number of defects, detectable 250 ms after the transition, decreases as a power law, in agreement with the prediction of the KZM and with the results reported in our previous work \cite{LamporesiKZM13}. Figure \ref{Figure6_alpha} shows how the power-law exponent varies with the transverse confinement at fixed axial confinement. 

The predictions of the KZM for a harmonically trapped condensate, with the critical exponents taken from the F-model, are $\alpha=7/6$ and $7/3$ in case of solitons and quantized vortices, respectively (see also Fig.~\ref{Figure6_alpha}). As outlined in Sec.~\ref{sec:theory}, the nature of the spontaneously created defects is strongly linked to the relative magnitude of $\Delta r$, $\hat{\xi}$ and $\xi_l$. 

At $T_{\mathrm{c}}$, large phase fluctuations are present and it is impossible to define defects for such early times \cite{Zurek09}. Let us consider the first small nucleus of the forming BEC, containing just about 1\% of the total atom number in the system at $T_{\mathrm{c}}$ (see also Fig. \ref{Figure3_BECTransition}). We can estimate $\Delta r$ as twice the transverse Thomas--Fermi radius $R_{\mathrm{rad}}$. In the various experimental conditions reported here, $\Delta r$ would then range between 10 and 20 $\mu$m and $\xi_l$ would be already smaller than 1 $\mu$m. As the system is further cooled, $\Delta r$ grows while $\xi_l$ becomes smaller and smaller, 
and hence we always have $\Delta r \gg \xi_l$. Finally, in order to estimate the average domain size $\hat{\xi}$ according to Eq.~(\ref{eq:kzm-xi-hat}), we can proceed as follows. Let us assume the parameter $\xi_0$ to be of the order of $\lambda_{dB}=\hbar/\sqrt{2\pi m k_{\mathrm{B}} T_{\mathrm{c}}}$ and $\tau_0$  of the order of the collisional time at center of the sample, where the BEC is nucleated, $\tau_0\simeq (\rho_0 \sigma_{\mathrm{coll}} v_{\mathrm{th}})^{-1}$, with $\sigma_{\mathrm{coll}}=8\pi a^2$ and $v_{\mathrm{th}}=4\sqrt{k_{\mathrm{B}} T_{\mathrm{c}}/(\pi m)}$. With these numbers at hand, one obtains $\hat{\xi}$ ranging from 1 to 3 $\mu$m, hence about one order of magnitude smaller than $\Delta r$. This would suggest that in our experimental conditions, the formation of vortices is always favored. However, the possible role of the multiplicative $f$ factor (from $1$ to $10$) mentioned in Sec.~\ref{sec:theory} makes it not easy to draw definite conclusions. 
For comparison, a direct measurement of $\hat{\xi}\simeq 1\ \mu$m was performed on a uniform Bose gas in Ref. \cite{Navon15}.

The data in Fig.~\ref{Figure6_alpha} are consistent with the nucleation of vortices for the two largest values of $\omega_{\mathrm{rad}}$, while they are closer to the soliton formation for smaller $\omega_{\mathrm{rad}}$. Such a deviation cannot be easily explained by the above analysis in terms of  $\Delta r$, $\hat{\xi}$ and $\xi_l$, which would rather suggest an opposite trend. However, the comparison with theory must be taken with care. On the one hand, the experimental error bars are still too large to make definitive statements. On the other hand, the KZM predictions for $\alpha$ assume a spatially uniform temperature profile in the system during the quench, while the temperature profile is actually nonuniform along the axis of our elongated condensate due to the different thermalization times in the axial and transverse directions. A derivation of the KZ exponents when the system exhibits inhomogeneities in both the density and temperature profiles is not yet available and might explain the variations of $\alpha$ with $\omega_{\mathrm{rad}}$.    

Finally, it is worth recalling that in our previous work~\cite{LamporesiKZM13} we presented a measurement of $\langle N_{\mathrm{d}} \rangle$ similar to the one of Fig.~\ref{Figure6_DefectScaling}, for an aspect-ratio of $10$. In that case we found $\alpha=\num{1.38\pm0.06}$ which is fully consistent with the new data of the present work, despite several differences in the experimental procedures.

\subsection{Small $\tau_{\mathrm{Q}}$: Saturation of defect number  \label{sec:saturation}}

According to the KZM the defect number in the system should follow a power-law scaling for all $\tau_{\mathrm{Q}}$.
Figure~\ref{Figure6_DefectScaling} shows, instead, that for fast quenches, $\langle N_{\mathrm{d}}\rangle$ clearly saturates. The values of $N_{\mathrm{sat}}$ resulting from the fits to the four datasets are reported in Table \ref{tab:fit-results} and their average value is $2.4(3)$. Here we suggest that such a saturation, which is almost insensitive to the change of the radial confinement, might originate from the post-quench dynamics of the condensate. Indeed, one must keep in mind that, for the data in Fig.~\ref{Figure6_DefectScaling}, the counting of defects is performed $250$~ms after the BEC transition. 

\begin{figure}[!t]
\includegraphics[width=\columnwidth]{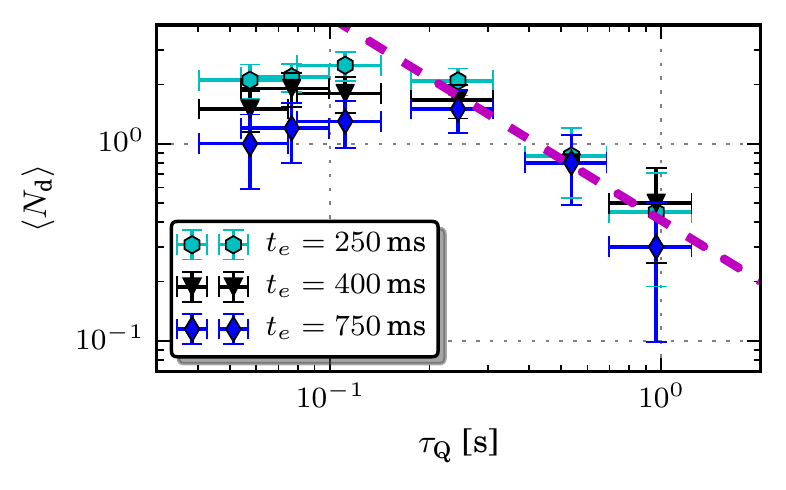}%
\caption{\label{Figure8_te} Average number of defects as a function of the quench time with a fixed $\omega_{\mathrm{rad}}=2\pi\times131$ Hz, for different evolution times. We note that the evolution time does not  substantially influence the linear behavior, which can be fitted with a single power-law (dashed purple).}
\end{figure}

The saturation for fast quenches was not observed in Ref.~\cite{LamporesiKZM13} because such high evaporation rates were not investigated. However, Refs.~\cite{LamporesiKZM13,Donadello14,Serafini15} report signatures of a post-quench dynamics and a finite lifetime of defects in elongated BECs as the ones studied here. Such a phenomenon, not considered in the KZM, may likely alter the defect counting, at least in the fast quenches regime. 

In order to investigate the effect of the condensate dynamics on our measurements, we repeat the whole set of measurements of $\langle N_{\mathrm{d}} \rangle$, in the case of $AR=10.1$,  for different values of the evolution time: $t_{\mathrm{e}}=250, 400$ and $750$~ms.
The results are reported in Fig.~\ref{Figure8_te}. We first observe that, in the power-law regime for large $\tau_{\mathrm{Q}}$, the data mostly overlap and the power-law exponent $\alpha$ looks insensitive to the evolution time. Instead, in the saturation region, the observed $\langle N_{\mathrm{d}} \rangle$ exhibits a clear dependence on $t_{\mathrm{e}}$: for longer evolution times the saturation occurs at lower defect numbers, suggesting a nonnegligible role of the vortex-vortex interaction that might enhance the vortex number decay.

\begin{figure}[b!]
	\centering
	\includegraphics[width=\columnwidth]{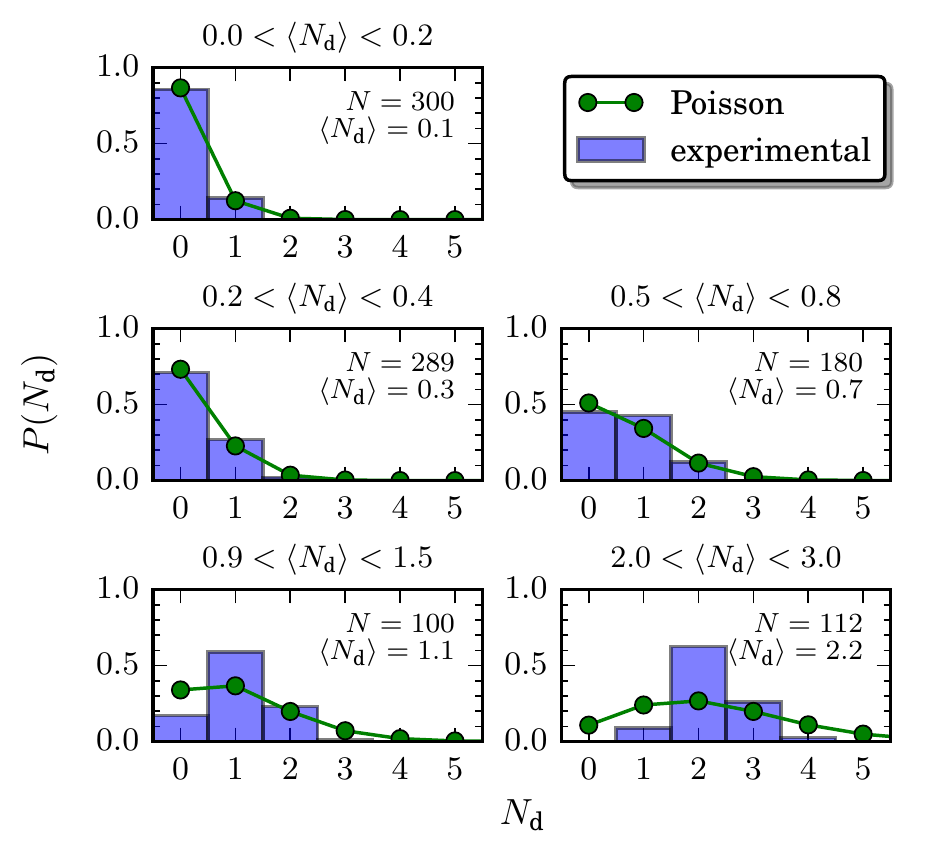}
	\caption{ Defect counting statistics. Histograms show the measured occurrence probability of $N_{\mathrm{d}}$ for given intervals of $\langle N_{\mathrm{d}} \rangle$ in the data reported in Fig.~\ref{Figure6_DefectScaling}. The number of experimental runs $N$ considered for each histogram is reported. Histograms are compared to the Poisson distribution which has its mean value equal to the central value of the considered bin. The agreement with the experimental data is good for small values of $\langle N_{\mathrm{d}} \rangle$, i.e., in the power-law regime (upper panels), while it is bad in the saturation regime (lowest two panels).}
	\label{Figure9_Poisson}
\end{figure}

Using the data reported in Fig.~\ref{Figure6_DefectScaling} and Fig.~\ref{Figure8_te} and assuming a Poisson distribution of $N_{\mathrm{d}}$, given the stochastic nature of the KZM, we try to reconstruct the amount of defects created during the quench, before they start interacting and decaying in number. To this purpose we proceed as follows. We first bin the experimental data of Fig.~\ref{Figure6_DefectScaling} by grouping the points lying in given intervals of $\langle N_{\mathrm{d}} \rangle$, independently of $\tau_{\mathrm{Q}}$ and $\omega_{\mathrm{rad}}$. For each interval of $\langle N_{\mathrm{d}} \rangle$ we plot the histogram of the measured occurrence probability of a given $N_{\mathrm{d}}$ in $N$ experimental runs and compare it to the Poisson distribution which has its mean value at the center of the considered bin, as shown in Fig.~\ref{Figure9_Poisson}. As one can see, for cases corresponding to the power-law regime, where $\langle N_{\mathrm{d}} \rangle \lesssim  1$, there is a good agreement between the experimental data and the Poisson distribution, while the distributions in the saturation regime, where $\langle N_{\mathrm{d}} \rangle \gtrsim 1$, show clear deviations. Of course, the distribution of defects observed after the evolution time is not the distribution which would be observed just after crossing the BEC transition, which would be Poissonian with a larger $\langle N_{\mathrm{d}} \rangle$. However, if the decay time of each single defect is independent of the actual number of defects present in the same condensate, then the overall effect of the decay would be a decreasing of $\langle N_{\mathrm{d}} \rangle$ with $t_{\mathrm{e}}$ independent of the quench rate, but keeping a Poissonian distribution. Conversely, if the decay time of the single defect  depends on the presence of other defects in the condensate, due to mutual interactions, then $\langle N_{\mathrm{d}} \rangle$ would decrease with $t_{\mathrm{e}}$ differently for different quench rates, thus producing a non-Poissonian distribution of the observed defects.  

\begin{figure}[t!]
	\centering
	\includegraphics[width=\columnwidth]{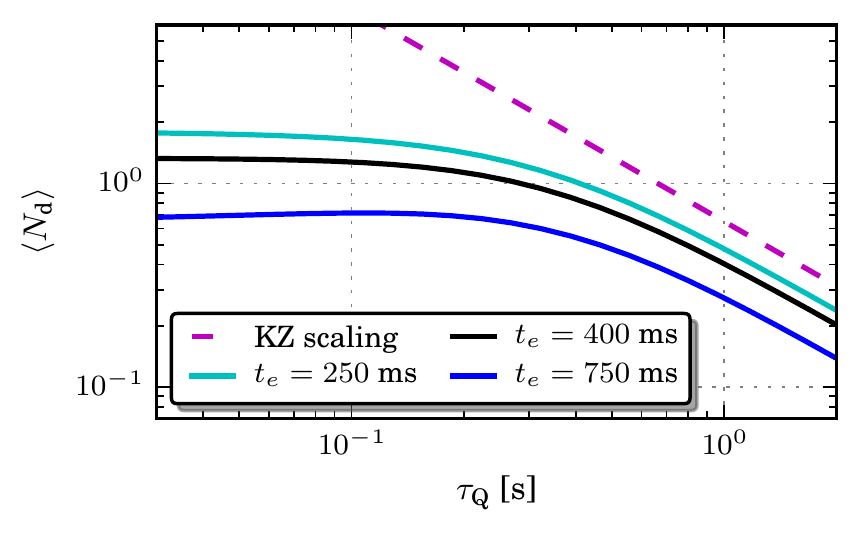}
	\caption{Predictions for the average defect number {\it versus} quench time, including the effects of defect decay during the evolution time after the BEC transition (see text).  The dashed line is the KZ power-law scaling (up to an arbitrary vertical offset) with the same exponent $\alpha$  of the experimental data of Fig.~\ref{Figure8_te}. The solid curves, from top to bottom, correspond to the average defect numbers expected after an evolution time $t_e=250, 400$ and $750$~ms, respectively. }
	\label{Figure10_Simulation}
\end{figure}

In Ref.~\cite{Serafini15} we indeed found that the life time of a vortex is the same if a condensate has one ($\tau_1$, of the order of $1\ $s) or two ($\tau_2 \simeq \tau_1$) vortices, but it is shorter if the vortices are three ($\tau_3$, of the order of $0.5\ $s) or more. We can now use this information in combination with Figs.~\ref{Figure8_te} and \ref{Figure9_Poisson}. Let us start from the slow quench regime where the majority of the condensates host very few vortices. In this regime, we can  assume that the post-quench dynamics  produces just a vertical shift of $\langle N_{\mathrm{d}} \rangle$, without changing the power-law scaling. This allows us to infer what would be the value of $\langle N_{\mathrm{d}} \rangle$, even for faster quenches, if the defects were observed just after the transition and under the assumption that the KZ scaling law is valid in the whole range of $\tau_{\mathrm{Q}}$: in fact, it is enough to extrapolate the observed power law to the whole range of $\tau_{\mathrm{Q}}$ (purple dashed line in Fig.~\ref{Figure8_te}), adding an upward vertical shift as a free parameter. For each $\tau_{\mathrm{Q}}$,  we calculate the Poisson distribution of defects corresponding to such extrapolated value $\langle N_{\mathrm{d}} \rangle$. Then, the occurrence $P(N_{\mathrm{d}})$ of each defect number $N_{\mathrm{d}}$ in the distribution is reduced by using an exponential decay, $\exp(-t_{\mathrm{e}}/\tau_{\mathrm{d}})$, with the life times $\tau_1$, $\tau_2$ and $\tau_3$ taken from Ref.~\cite{Serafini15}. We also assume that $\tau_{\mathrm{d}}$, with d$=4$ or larger, is much shorter than any other relevant timescales (in practice we truncate the Poissonian distribution for $N_{\mathrm{d}}\geq4$ and then renormalize it before applying the time evolution). In this way, the initial Poissonian distribution, at a given $\tau_{\mathrm{Q}}$, is deformed and the average number of defects after the evolution time $t_{\mathrm{e}}$ is lowered by an amount which depends on the quench rate. This effect is larger for fast quenches, because the initial number of defects is larger, thus making their decay faster. 

Using this extrapolation protocol, we obtain a prediction for the average defect number which would be observed after an evolution time $t_{\mathrm{e}}$ starting from a Poissonian distribution fixed by the KZ scaling. The resulting curves are shown in Fig.~\ref{Figure10_Simulation}. The arbitrary upward shift of the dashed line has been adjusted in such a way that the predicted values of $\langle N_{\mathrm{d}} \rangle$ at large $\tau_{\mathrm{Q}}$ are close to the observed ones. The overall qualitative behavior of the curves predicted by this simple model is rather similar to the experimental observations in Fig~\ref{Figure8_te}:  a clear saturation for fast quenches emerges and  the time scale of the transition between the two regimes falls in the same range of $\tau_{\mathrm{Q}}$. Despite the strong assumptions made, which would require extensive measurements and simulations to be validated, this analysis suggests indeed that the defect lifetime can be a possible explanation for the saturation that we observe in the KZ scaling.

\section{Conclusions\label{sec:conclusions}}

In this work, we measure the number of defects spontaneously created in a BEC after cooling a trapped bosonic gas of sodium atoms across $T_{\mathrm{c}}$ with different quench rates and for several transverse confinements. We clearly distinguish two regimes: 
a) For slow cooling rates (large $\tau_{\mathrm{Q}}$) a power-law behavior of the average defect number is observed as predicted by the KZM. In the case of strong confinement, our results are consistent with the exponent predicted for a harmonically trapped elongated gas in which vortices are spontaneously produced. On the other hand, in the case of weaker confinement, the exponent turns out to be slightly smaller.
b) For fast cooling rates (small $\tau_{\mathrm{Q}}$) we see a clear saturation of the measured average defect number to a value around $2.4$, almost independent of the transverse confinement. We provide a qualitative interpretation in terms of the post-quench dynamics and interaction between vortices.

These results, which extend and improve our previous observations of Ref.~\cite{LamporesiKZM13}, represent a further step toward a better understanding of the KZM in inhomogeneous bosonic systems. They can also stimulate the investigation of the dynamics of quantized vortices in quenched superfluid with boundaries. A possible approach consists of performing extensive numerical simulations of the condensate dynamics at finite temperature using, for instance, the stochastic Gross-Pitaevskii equation in conditions similar to that of our system. Work in this direction is in progress \cite{Liu16}.

\begin{acknowledgments}
We are grateful to N.P.~Proukakis and I-Kang~Liu for sharing preliminary results of their 3D SGPE simulations and for many discussions. We thank W.~Zurek, A.~del~Campo, Z.~Hadzibabic, G.~Morigi, T.~Calarco, M.~Tylutki and F.~Larcher for insightful discussions. T.B. acknowledges the EU QUIC project for fundings. This work was financially supported by Provincia Autonoma di Trento.
\end{acknowledgments}

\bibliography{bibliography}

\end{document}